\documentclass[12pt,preprint]{aastex}

\shorttitle{Oxygen and magnesium evolution}
\shortauthors{McWilliam et al.}

\begin{document}


\title{The Evolution of Oxygen and Magnesium in the Bulge and Disk of the Milky Way
   }


\author{A. McWilliam\altaffilmark{1}}
\author{F. Matteucci\altaffilmark{2,3}}
 \and
\author{S. Ballero \altaffilmark{2,3}}
\author{R.M. Rich\altaffilmark{4}}
\author{J.P. Fulbright\altaffilmark{5}}
\and
\author{G. Cescutti\altaffilmark{2}}



\altaffiltext{1}{Observatories of the Carnegie Institution of Washington, 
     813 Santa Barbara St., Pasadena, CA 91101}
\altaffiltext{2}{Dept. of Astronomy, University of Trieste,
      Via G.B. Tiepolo, 11, 34100 Trieste, Italy}
\altaffiltext{3}{I.N.A.F.- Osservatorio Astronomico of Trieste,
      Via G.B. Tiepolo 11, 34100 Trieste, Italy}
\altaffiltext{4}{Dept. of Physics and Astronomy,
      UCLA, 430 Portola Plaza, Box 951547, Los Angeles, CA 90095-1547}
\altaffiltext{5}{Department of Physics and Astronomy,
    Johns Hopkins University, Baltimore, MD 21218}


\begin{abstract}
We show that the Galactic bulge and disk share a similar, strong, decline in
[O/Mg] ratio with [Mg/H].  The similarity of the [O/Mg]
trend in these two, markedly different, populations suggests that the strong decline is due
to a metallicity dependent modulation of the stellar yields
%
from massive stars, by mass loss from winds, and related to
the Wolf-Rayet phenomenon, as proposed by McWilliam \& Rich (2004).

We have modified existing models for the chemical evolution of the Galactic bulge and the 
solar neighborhood with the inclusion of metallicity-dependent oxygen yields from theoretical
predictions for massive stars that include mass loss by stellar winds.  Our results 
significantly improve the agreement between predicted and observed [O/Mg] ratios in the 
bulge and disk above solar metallicity; however, a small zero-point normalization problem remains
to be resolved.   The zero-point shift indicates that either the semi-empirical yields of
Fran\c cois et al. (2004) need adjustment, or that the bulge IMF is not quite as flat
as found by Ballero et al. (2007ab); the former explanation is preferred.

Our result removes a previous inconsistency between the interpretation of [O/Fe] and 
[Mg/Fe] ratios in the bulge, and confirms the conclusion that the bulge formed more rapidly
than the disk, based on the over-abundances of elements produced by massive stars.  We also 
provide an explanation for the long-standing difference between [Mg/Fe] and [O/Fe] trends 
among disk stars more metal-rich than the sun.
\end{abstract}





\section{Introduction}

Oxygen and magnesium are so-called alpha elements, thought to be produced in the cores of
massive stars during the hydrostatic phase.  They are among the elements most easily 
understood because their synthesis is restricted to well known processes in the cores of
massive stars and they are thought to be unaffected by the complications of explosive 
nucleosynthesis (e.g. Woosley \& Weaver 1995; henceforth WW95); thus, one might expect that
O and Mg abundances should vary in lock-step in all, or most situations.  O and Mg are
particularly interesting as probes of massive stars and the star formation rate in chemically
evolving stellar systems.  A classic example of this is the paper by Matteucci \& Brocato 
(1990) that predicted enhancements of alpha elements in the Galactic bulge and giant 
elliptical galaxies, relative to the
solar neighborhood, due to the high star formation rate expected in the bulge and
ellipticals.  Matteucci \& Brocato (1990) also predicted deficiencies of alpha elements in
galaxies thought to have low star-formation rates, such as the LMC.  The simple reason for this is
that in high SFR systems high metallicities are reached before Type~Ia supernovae can
inject significant amounts of Fe to the ISM; thus, the alpha/Fe ratios remain at the high
levels expected to be produced by core-collapse, Type~II, supernovae.

Early observational confirmation of the predictions of Matteucci and Brocato (1990) came
from McWilliam \& Rich (1994) for the bulge and Shetrone, C\^ot\'e, \& Sargent (2001) for 
dwarf spheroidal galaxies.  These early results have been verified and improved upon by 
many others for the bulge, including Rich \& McWilliam (2000), McWilliam \& Rich (2004),
Fulbright, McWilliam \& Rich (2006, 2007), Zoccali et al. (2006), 
Lecureur et al. (2007) and also for dwarf galaxies (e.g. Hill 1997; Geisler et al.  2005;
Venn et al. 2004; Tolstoy et al. 2003; McWilliam \& Smecker-Hane 2005).

While McWilliam \& Rich (1994) found alpha enhancements in Galactic bulge stars, a they noted that 
not all alpha elements were enhanced.  McWilliam \& Rich (1994)
proposed that a bulge IMF skewed toward massive stars could explain this selective alpha
enhancement, based on the mass-dependent yields of WW95, that indicated  relatively greater production
of Mg than Ca and Si from higher mass Type~II SNe.  McWilliam \& Rich (2004) found a
steeply declining [O/Fe] ratio  in the bulge stars with a slope almost consistent with no 
production of O above [Fe/H] $>$ $-$0.5. 
They suggested an O yield decline due to winds from massive stars, related to the WR phenomenon.
Later Fulbright et al. (2007)  noted the apparent discord between nearly flat, 
enhanced, [Mg/Fe] and yet declining, although enhanced, [O/Fe] in the bulge above 
[Fe/H] $\sim$ $-$1.0; again, they suggested that declining metallicity dependent O yields 
related to the WR phenomenon.  More recently, Lecureur et al. (2007) also found a decreasing
trend for [O/Mg] in the bulge stars, and suggested a difference in the stellar yields of
O and Mg from massive stars.

This observed difference between Mg and O trends could potentially lead to a conflict
in the estimated star formation rate and formation timescale of the bulge: if the steeper decline of
[O/Fe], compared to [Mg/Fe], with increasing metallicity is due to the addition of Fe from Type~Ia SNe,
then a lower SFR and longer bulge formation timescale would be indicated from O than for Mg.  If the
different decline rates for O and Mg are due to the supernova yields it is necessary to understand them 
in a consistent picture of nucleosynthesis and chemical evolution.

Different slopes in the [Mg/Fe] and [O/Fe] ratios are visible in metal rich stars in the 
solar vicinity, where the [Mg/Fe] seems to reach a plateau for [Fe/H] $>$ 0, but the 
[O/Fe] ratio continues to decrease (see Edvardsson et al.1993; Castro et al. 1997;
Feltzing \& Gustafsson 1998; Bensby et al. 2005). 
This trend also suggests the effect of strongly metal dependent O yields in 
metal rich massive stars.

Stellar yields can be affected by mass loss in massive stars.  Maeder(1992) 
computed the wind contributions in He, C, N, O and Ne in stars in the mass range 
1-120$M_{\odot}$ and for metallicities Z=0.001 and Z=0.02.  He found that at metallicities 
of solar, and above, large amounts of He and C are ejected into the interstellar
medium, reducing the synthesis of heavier elements; thus C and He are produced at the
expense of O.  The
same effect is not seen at low metallicities and therefore the nucleosynthesis production 
very much depends on the initial stellar metallicity.  The metallicity dependence of the
stellar wind is due to
the opacity of the material in the radiation-driven mass loss of massive stars,
as originally proposed by Lucy \& Solomon (1970).
The effect of mass loss on stellar yields as a function of metallicity is particularly 
strong for initial stellar masses $> 25-30M_{\odot}$: those which will become Wolf-Rayet 
(WR) stars and eventually explode as Type Ib/c supernovae.  By
means of these stellar models Maeder was able to reproduce the number statistics of WR and
O-type stars in nearby galaxies with various metallicities.



Later, Langer \& Henkel (1995) also computed He and CNO isotope yields for mass-losing stars 
in the mass range $15\le M/M_{\odot} \le 50$.  They also found drastic differences 
relative to models without mass mass loss for the yields of oxygen  in metal-rich stars 
more massive than $30M_{\odot}$.  

More recently, Meynet \& Maeder (2002, 2003, 2005)  have computed  a grid of models for 
stars with masses $>20M_{\odot}$, including rotation and metallicity dependent mass loss.
The effect of metallicity dependent mass loss in decreasing the O production in massive 
stars was confirmed, although they employed significantly lower mass loss rates, based on 
work by Vink et al. (2000) and Nugis \& Lamers (2000).  The new and improved mass-loss 
rates are factors of 2 to 3 lower than previously adopted by Maeder (1992); however, with 
these mass-loss rates and the inclusion of rotation the models are able to reproduce the 
frequency of WR/O stars, the observed WN/WC ratio, and the observed ratio of 
Type~Ib/Type~Ic supernovae at different metallicities. It appears that the earlier high 
mass-loss rates made-up for the omission of rotation in the stellar models.
In galactic chemical evolution models, the effect of the Maeder (1992) yields was
studied by Prantzos, Aubert \& Audouze (1996), who concluded that the mass loss in
massive stars has strong effects on the production of C and O.

In summary, the effect of metallicity dependent mass loss
on stellar yields appears relevant for stars with solar metallicities or larger, and 
therefore it should be taken into account in computing the bulge chemical evolution and the
late stages of chemical evolution in the solar vicinity.

In this work, we present new results for the chemical evolution of the galactic 
bulge and the solar vicinity concerning the evolution of O, Mg and Fe.  We adopt
detailed models developed for the bulge (Ballero et al. 2007a) and for the solar 
neighborhood (Chiappini et al. 1997) in which we include metallicity dependent yields for
O from WW95 up to solar metallicities and the mass loss dependent 
yields for O from Maeder (1992) for metallicities above solar. For the yields of Fe and Mg,
which are unaffected by mass loss, we adopt those of WW95 with corrections 
suggested by Fran\c cois et al. (2004) for Mg, in order to obtain a very good fit of the solar 
vicinity abundance patterns.  By means of these models we will predict the behavior of the
[O/Mg] ratio both in the bulge and in the solar neighborhood to test whether the mass loss
O yields can solve the problem.

The paper is organized as follows: in Section 2 a brief description of the chemical 
evolution models is given, in Section 3 \& 4 the theoretical results are compared with the data
and in Section 5 some conclusions are drawn.

\section{Observational Evidence for Metallicity-Dependent Oxygen Yields}

The difference between the trends of [Mg/Fe] and [O/Fe] seen in the Galactic bulge
found by McWilliam \& Rich (2004), Fulbright et al. (2007) and confirmed by
Lecureur et al. (2007), adds confusion to the interpretation of the evolution timescale
for the bulge, because oxygen declines precipitously, while Mg changes very slowly, with 
metallicity.  One might have expected the two abundance trends to be very similar, since O
and Mg are both thought to be produced only by stars that end as Type~II supernovae 
(e.g. WW95); however, the Galactic bulge observations clearly show that the abundances of
O and Mg do not vary in lockstep.  Indeed, the dissimilarity of the bulge [O/Fe] and 
[Mg/Fe] trends was the reason why McWilliam \& Rich (2004) suggested a metallicity-dependent
decline in oxygen yields, related to the Wolf-Rayet phenomenon.

Because O and Mg are produced only in the hydrostatic cores of massive stars, then if the
O/Mg yield ratio declines as a function of metallicity, the same trend should be present in
stellar systems, no matter what the star formation rate, providing that the formation
timescale is long enough to permit all masses of Type~II SNe to occur.   To test the
metallicity-dependence of the O/Mg ratio in this paper we compare the abundance ratio of O/Mg 
seen in the Galactic bulge and the Galactic thin and thick disks.  In particular we use [Mg/H] and 
[O/H] as metallicity indicators, rather than [Fe/H], in order to eliminate the effect of Fe
from Type~Ia supernovae.

The observational evidence for a decrease in the oxygen yield, relative to magnesium,
from Type~II supernovae with increasing metallicity is summarized in Figures~1a and 1b.  
In Figure 1a we show a plot of the [O/Mg] versus [Mg/H] for the bulge from 
Fulbright et al. (2007), Origlia \& Rich (2002), Rich \& Origlia (2005), Zoccali et al. (2004),
and Lecureur et al. (2007), compared with points for the solar neighborhood thin and thick 
disks from Bensby et al. (2005).  Contrary to what one might expect for two alpha-elements 
produced by the same stars the [O/Mg] ratio is not flat, but declines steeply for
[Mg/H] values larger than $-$0.5 dex.  The same effect is visible both in the 
bulge and disk stars.   
Note that in Figure~1b we show the same observed [O/Mg] ratio, but with [O/H] as the
metallicity indicator.  For the bulge metallicity points we employ
the mean [O/H], based on [Fe/H] and a fit to the trend of [O/H] versus [Fe/H]; it
was motivated to reduce the scatter in the [O/H] metallicity axis resulting from
the use of only the 6300\AA\ [O~I] line for the oxygen abundances.  Oxygen is useful
as a metallicity indicator, despite the noise, because it contributes more than half of the
total metallicity, Z.
We have avoided using the recent oxygen abundances for thick disk stars by Reddy, Lambert
\& Allende Prieto (2006),
because these were based on the high excitation O~I triplet lines near 7771\AA , and not
the robust [O~I] 6300\AA\ indicator; the high excitation O~I lines require
correction for non-LTE effects that are somewhat uncertain.


In both figures there is a clear decline in the bulge [O/Mg] ratio of $\sim$0.8 dex from 
low to high metallicity, including a markedly steeper descent above solar metallicity.  It 
is notable, and very important, that the [O/Mg] trends for the Galactic disk closely 
overlap the bulge trends, and that both show the steeper decline above solar metallicity.  
A similarity in the evolution of the products of Type~II supernovae in the disk and
bulge is expected because O and Mg are produced on the same timescales.  On the other
hand, we expect to see the effect of the time delay model in the [$\alpha$/Fe] ratios.

We suggest that, to within the measurement uncertainties, the disk and bulge [O/Mg] 
versus [Mg/H] trends are the same.  However, there is some leeway in this interpretation
with a possibility that the bulge [O/Mg] ratios slightly exceed the disk for a given
[Mg/H].  Some of the individual bulge studies may show offsets from the disk trend, but
these are small; in this regard we note that in the Fulbright et al. (2007) study there was
some evidence to support a systematic downward zero-point correction to the oxygen abundances, 
by $\sim$0.04 dex; if applied this would make a slight improvement to the already good
agreement between disk and bulge O/Mg relations.

Figures 1a and 1b compare the composition of the bulge, formed within the
initial $\sim$1 Gyr after the Big Bang, with the composition of disk stars formed up
to the present epoch.  We note that the trend with [Mg/H] as metallicity indicator 
is particularly important because the [Mg/Fe] ratios in the bulge are significantly 
higher than in the disk due to Fe from Type~Ia supernovae.  Given these differences 
between the disk and bulge it is 
remarkable that the [O/Mg] trends with [Mg/H] and [O/H] follow each other so closely.  

We begin to understand the O/Mg ratio trend by appreciating that both O and Mg are thought
to be produced only by core-collapse supernovae.  In particular it should be remembered
that O production increases strongly with pre-supernova mass (e.g. WW95),
but more important for this investigation is the trend of O/Mg yield ratio 
with supernova progenitor mass.
Following recent works using our adopted chemical evolution model (e.g. Ballero et al. 2007ab)
the Mg yields employed here are the semi-empirical values of Fran\c cois et al. (2004), found
by inverting a chemical evolution model to match the composition of stars in the solar
neighborhood; also following these papers we take the O yields from WW95, with the yield
beyond the WW95 mass range fixed at the value for the 40$M_{\odot}$ model.  These yields show
a strongly increasing O/Mg ratio toward higher supernova progenitor masses, with a range 
exceeding 1.5 dex.  This high sensitivity to progenitor mass is similar to the predictions for 
hypernovae computed by Fryer, Young \& Hungerford (2006); however, results from a host 
of other calculations including: WW95, Tsujimoto et al. (1995; the same as Nomoto et al.
1997), Limongi \& Chieffi (2003) and Kobayashi et al. (2006) show very shallow increases,
or nearly flat, O/Mg trends with mass.  One of the steeper predicted O/Mg yield ratios
with mass, from early works, is due to Arnett (1991), with a range of about 0.5 dex.
We note most of the theoretical predictions are for masses below about 40M$_{\odot}$,
although Tsujimoto et al. (1995) go to 70M$_{\odot}$.

If the O/Mg yield ratio is strongly sensitive to the mixture of supernova masses,
the observed abundance ratios will be affected by the initial mass function (IMF). 
In particular, a flatter bulge IMF than the galactic disk would produce a higher
[O/Mg] ratio in the bulge, if other parameters remain the same.  However, the data seems
to show a large overlapping of the [O/Mg] ratio in the bulge and disk stars, although 
some bulge stars show a higher ratio.   However,
Ballero et al. (2007a) suggested a flat IMF for the bulge in line
similar to previous results (e.g. Matteucci \& Brocato 1990; Matteucci et al. 1999),
based on the observed stellar metallicity distribution for the bulge, which
cannot be reproduced with the same IMF as the disk with the current set of parameters.

Thus, the similarity of the O/Mg ratios in the bulge and disk suggest that either
the O/Mg yield ratio is quite insensitive to stellar mass, in which case the yields
of Fran\c cois et al. (2004) need to be modified, or that the IMF for the
bulge is similar to that of the disk, with much less of a difference in slope than
indicated by Ballero et al. (2007a).  Solutions might also be found by modifying both
the element yields and the bulge IMF slope.
More data  will help to decide whether the abundance ratios in the bulge are higher or
equal to those in the solar vicinity, while further theoretical and empirical studies 
may constrain the O and Mg yields from massive stars.

If O/Mg yields are a strong function of stellar mass the observed downward slope in the
O/Mg, with increasing [Mg/H], could be explained by a metallicity-dependent IMF, that 
increases the fraction of low mass supernovae at higher metallicity; this would 
have to apply to both the disk and the bulge, but there is no strong evidence of a variable 
IMF in the disk.  Therefore, we discard the notion of a metallicity-dependent IMF to explain 
the observed O/Mg trend.

We note that if O/Mg yields do increase strongly with stellar mass, then during initial
enrichment O/Mg ratios would be higher
than at later times, due to high oxygen yields from the massive stars that are the first
to end as supernovae.  Thus, over the time period spanning the lifetime of massive stars
there should be a decline in the average O/Mg ratio of a system.  This low-metallicity effect
might be seen at a higher metallicity in the bulge than other Galactic locations, due to the
high star formation rate in the bulge.

We adopt the simple notion that the observed similarity of the O/Mg slopes for both the 
Galactic disk and bulge reflects a metallicity modulation of the element yields from core 
collapse supernovae; in principle this could be due to O or Mg yields, or both.

The bulge results are similar to those of Bensby, Feltzing \& Lundstr\"om (2004), who identified the
decline in [O/Mg] versus [Mg/H]
in the Galactic thin and thick disk populations.  They considered the possibility that metallicity-dependent 
oxygen yields from massive stars could explain the observed trend, but they dismissed this idea
based on their interpretation of the oxygen yields of massive stars from models, including rotation, by
Meynet \& Maeder (2002).  Our understanding is that while rotation does increase the core mass, the
effect of decreased oxygen yields with increasing metallicity, due to mass loss, is clear in the
works of Meynet \& Maeder (2002, 2003, 2005); the effect on the yields is particularly noticeable
for the most massive stars above solar metallicity (Meynet \& Maeder 2005).  This result holds, despite 
the recent decrease in adopted mass-loss rates, compared to Maeder (1992).

\section{The chemical evolution models}

For the chemical evolution of the Milky Way we adopted the model of Chiappini et al. (1997)
with updated nucleosynthesis prescriptions as in Fran\c cois et al. (2004). This model is 
the so-called two-infall model where the halo and part of thick disk are formed during a 
first relatively short episode ($< $ 2 Gyr) of accretion and star formation, whereas the 
thin disk formed out of a second independent infall episode which lasted much longer (8 Gyr
in the solar vicinity) and formed the disk ``inside-out''. This model takes into account in
detail the stellar lifetimes, detailed nucleosynthesis and supernovae of all types. It 
follows the evolution in space and time of 35 chemical species.  The adopted IMF is the one
from Scalo (1986) and is considered constant in space and time.  The nucleosynthesis 
prescriptions are: for massive stars ($M>10M_{\odot}$) the basic yields are those from 
WW95 although for some elements, such as Mg, we made 
corrections following the suggestions of Fran\c cois et al. (2004). In particular, the 
yields from stars in the range 10-20$M_{\odot}$ were increased in order to fit the 
observed Mg absolute solar abundance.  The well-known problem of low Mg yields has been
discussed by others (e.g. Thomas, Greggio \& Bender, 1998).  The yields for Type Ia SNe,
which are assumed to originate from single-degenerate systems, are taken from Iwamoto et al.
(1999). Finally the yields for low and intermediate mass stars are those from van den 
Hoeck \& Groenewegen (1997).

The model for the bulge is that developed by Ballero et al. (2007a); exactly the same 
nucleosynthesis prescriptions and SN progenitor models are adopted here. The bulge and
solar vicinity differ in the formation timescale and star formation efficiencies:
for the bulge the timescale is $\tau_{B}=0.1$ Gyr, compared to the formation timescale of
the solar vicinity, at $\tau_{SV}=8$ Gyr.  The star formation efficiencies (star formation 
rate per unit mass of gas) for the bulge and solar vicinity are: $\nu_{B}=20 Gyr^{-1}$, and 
$\nu=1Gyr^{-1}$ respectively.  This means that the bulge is assumed to have formed extremely 
rapidly, during a burst of star formation.  Ballero et al. (2007a) showed, in agreement with 
previous papers, that the initial mass function (IMF) in the bulge should be flatter than 
in the rest of the disk, as required by the comparison with the metallicity distribution of
bulge stars.  In this work we adopt the best model of the Ballero et al. (2007a) paper with
the following IMF: $X=0.95$ for stars with $m > 1M_{\odot}$ and $x=0.33$ for stars with  
$m \le 1M_{\odot}$. 

It should be said, however, that the slope of the IMF for massive stars could be higher
than x=0.95 and still be consistent with the stellar metallicity distribution in the
bulge, but the slope should be at maximum x=1.35.
This model provided a very good fit of the stellar metallicity 
distribution and also a good fit of the [$\alpha$/Fe] vs. [Fe/H] patterns, predicting a 
long $\alpha$-enhanced plateau for both O and Mg, as observed, but it predicted a flatter 
than observed behavior of the [O/Fe] for [Fe/H] $>0$.

\section{Results}

In order to investigate the potential effect of mass loss on predicted O/Mg ratios
we decided to update our current model with the Maeder (1992) yields that take into account
the effect of mass-loss for massive stars as a function of metallicity.  We used the 
Maeder (1992) yields, rather than the latest results of Meynet \& Maeder (2005), in order 
to maintain consistency with the pre-existing model employing the WW95 yields; both WW95 and 
Maeder (1992) models ignore the effects of rotation, but both provide similar yields near 
solar metallicity.  In this way our calculation should show how the metallicity-dependent 
yields affect the current model.
In Figure~2 we can see the differences 
between the yields of oxygen of WW95 and Maeder (1992) for two different initial stellar 
metallicities. The effect of the metallicity-dependent mass loss is evident in the figure:
especially for stars with masses larger than 25$M_{\odot}$ and solar metallicity the
O production is strongly depressed due to mass loss.  In Figure~3 we show the same data as
in Figure~1, but we have superimposed the theoretical predictions for the bulge and disk. 
As one can see in both cases the slope of the [O/Mg] ratio is very well reproduced by the 
Maeder yields. In particular, both the observations and our predictions indicate a sudden
steepening of the slope above solar metallicity, with good agreement between the predicted
and observed slopes below and above the break-point.

The $\sim$0.2 dex zero-point offset between our predicted bulge [O/Mg] trend suggests
that either the yields of Fran\c cois et al. (2004) require 
adjustment, or that the adopted IMF of Ballero et al. (2007a) is too flat, or a combination
of these effects.  Normalization of the predictions to solar abundances is also an issue.
Given that the bulge metallicity function suggests a flatter IMF than the disk
(Ballero et al. 2007ab) and the difficulties involved with inverting chemical
evolution models to obtain element yields, we suspect that the problem most likely arises
in the adopted O/Mg yields of Fran\c cois et al. (2004).  As we noted earlier 
the O/Mg trend with supernova progenitor mass, suggested by the results of
Fran\c cois et al. (2004), is much steeper than most theoretical predictions; if this
steep slope is in error we would expect to see an artificial zero-point offset in the
predicted O/Mg ratio.

Fortunately, the zero-point problem
has little effect on the main conclusion of this paper, namely: diminished oxygen yields, due to 
mass-loss from high metallicity Type~II supernovae, are responsible for the observed 
declining [O/Mg] trend with increasing [Mg/H] in the Galactic bulge and disk.  This resolves
the apparent discord between the bulge trends of [Mg/Fe] and [O/Fe] with [Fe/H]; now both
are consistent with a high star formation rate in the bulge.  
Resolution of the zero-point issue,
and the implication for Galactic chemical evolution, is too involved for the present work; 
we intend to address this problem in a forthcoming paper.

Our model results have been normalized to the solar abundances predicted by the 
Milky Way model, which gives a very good fit to the Asplund et al (2005) solar abundances.  
It is worth recalling that Asplund et al. (2005) found a lower O abundance than Grevesse 
\& Sauval (1998), whereas there is little difference for the abundances of Mg and Fe.
In Figure~4 we show the same plot as in Figure~3, but only for the bulge and including more
data from other sources (e.g.  Origlia, Rich \& Castro 2002; Origlia \& Rich 2004;
Origlia et al. 2005;  Rich \& Origlia 2005; Zoccali et al. 2004, 2006); the agreement between
the slope predicted with Maeder yields and the data is again very good.  Again, the
predicted absolute values are higher than the data, but this depends on the normalization to 
the solar abundances. In
Ballero et al. (2007a) where the solar abundances were those of Grevesse \& Sauval (1998), 
the predicted [O/Mg] was lower, due to the higher O solar abundance suggested by Grevesse 
\& Sauval relative to Asplund et al.

In Figure~4 we also show the same model predictions normalized to the Grevesse \& Sauval 
(1998) solar abundances.  The important points are that the slopes and break-point in 
the [O/Mg] trends with metallicity, and that both Milky Way and bulge results 
are normalized to the same solar abundances. It is also important to note that the 
predicted [O/Mg] in the bulge is higher than in the Milky Way.  In our models this mainly 
depends on the assumed IMF in the bulge which is flatter than in the solar vicinity.
Such a flatter IMF seems to be unavoidable in order to reproduce the observed stellar 
metallicity distribution in the bulge, as extensively discussed in Ballero et al. (2007a) 
and  Ballero, Kroupa \& Matteucci (2007b).
The bulge parameters determined by Ballero et al. (2007a) for the IMF, star
formation rate and infall timescale are not affected by the consideration of mass-loss 
dependent yields investigated here.  For the IMF and star formation rate 
the Ballero et al. (2007a) work was entirely constrained by [O/Fe] at metallicities
below solar, for which mass-loss dependent yields are not a factor.  While the [O/Fe] 
predictions of Ballero et al (2007a) provided a slightly better fit with in infall timescale 
near 0.7 Gyr$^{-1}$, the metallicity distribution function excluded values larger than 
0.1 Gyr$^{-1}$.  However, the metallicity-dependent oxygen yields considered here are
consistent with an infall timescale near 0.1 Gyr$^{-1}$, in agreement the value adopted by
Ballero et al.  (2007a).

It is also worth noting that, in Figure 3, our chemical evolution
model for the solar vicinity does not predict such high [Mg/H] values as observed by 
Bensby et al. (2005).  This is not a critical issue, since it may depend on many model 
parameters such as the present time infall rate, which could have been slightly 
overestimated, or the efficiency of star formation, which could have been slightly 
underestimated.  The important fact is that with the same nucleosynthesis prescriptions we 
reproduce the slope and break of the [O/Mg] ratio trend, thus supporting the original 
suggestion of McWilliam \& Rich (2004) and Fulbright et al. (2007).  

It is important to
consider that [O/Mg] versus [Mg/H] or [O/H] plots do not contain the effects produced by 
the chemical enrichment of Type~Ia SNe, but depend only on the different yields for O 
and Mg because these two elements are produced on similar timescales.
On the other hand, the effect of Type Ia SN enrichment is clearly visible in the 
[$\alpha$/Fe] versus [Fe/H] plots.  In these plots  the change in slope of the 
[O,Mg/Fe] ratios is mainly due to the time-delay with which the bulk of Fe is injected into
the ISM by Type~Ia SNe. This well known effect has been already extensively studied and 
commented in several previous papers (Matteucci \& Brocato, 1990; Matteucci, 
Romano \& Molaro, 1999, Ballero et al. 2007a). In particular, the history of 
star formation plays an important role in the [$\alpha$/Fe] versus [Fe/H] diagrams: in the 
bulge, which is assumed to have suffered an intense, burst-like, star formation we expect
a long plateau of enhanced-$\alpha$'s and a turning point occurring at [Fe/H]$\ge$ 0, as 
opposed to the Milky Way where the star formation proceeded much more smoothly and 
therefore we expect a turning point occurring at lower metallicities. The situation is even
more extreme in the dwarf galaxies (see Lanfranchi \& Matteucci, 2004) where the star formation
proceeded very slowly. In this case the turning point is expected at even lower metallicities,
thus having low [$\alpha$/Fe] ratios at low [Fe/H].
The predictions for  [O/Fe] versus [Fe/H] in the bulge compared with data can be seen in 
Figure 5, and the fit is again excellent.  In Ballero et al. (2007a) this plot was also shown
with the difference that the predicted [O/Fe] at high metallicities was flatter and in less 
good agreement with the observed points.

According to the Maeder (1992) prescription, and the later works of Meynet \& Maeder (2002, 
2003, 2005), our model clearly suggests that carbon should increase
in the bulge at metallicities above solar, due to the effect of mass-loss on 
massive stars.   We expect that the carbon enhancements could be significant, although we
have not performed detailed calculations at this point.  We shall investigate this 
qualitative prediction for the bulge composition in a forthcoming paper.

The metallicity-dependent decline in oxygen yield brings understanding to the long-standing
apparent discrepancy (Pagel private communication) between the trend of [O/Fe] at high [Fe/H]
and the [$\alpha$/Fe] trends for Mg, Si, Ca and Ti in the Galactic disk.  In the standard 
time-delay explanation for the decline of [$\alpha$/Fe] with [Fe/H] all these elements should 
show similar declines.  However, in the Galactic disk [O/Fe] continues a roughly linear 
decline above solar [Fe/H] to negative [O/Fe] values, as seen in the results of 
Edvardsson et al. (1993), Castro et al. (1997), Feltzing \& Gustafsson (1998), and 
Bensby et al. (2004). However, the other alpha elements reach a low plateau in [$\alpha$/Fe], 
at the solar ratio, beginning near [Fe/H]$\sim$ $-$0.4 to $-$0.2 dex; and this plateau 
continues to the most metal-rich disk stars.  The unusual behavior of oxygen is unexpected 
from the pure time-delay explanation for the decline in [$\alpha$/Fe] with increasing [Fe/H] 
in the Galactic disk; however, the time delay model very well describes the plateau seen in 
the other alpha elements, the alpha enhancements seen in bulge stars (e.g. McWilliam \& Rich 
1994; Fulbright et al. 2007), 
subsolar [$\alpha$/Fe] ratios seen in small subgroups of halo stars (e.g. Brown, 
Wallerstein \& Zucker 1997; Nissen \& Schuster 1997) and the decline in [$\alpha$/Fe] ratios 
at lower [Fe/H] in dwarf galaxies (e.g. Geisler et al. 2005).  The solution is that in the 
disk both the time delay reduction in [$\alpha$/Fe] and metallicity-dependent decline in
oxygen yields operate; this is why the [O/Fe] ratios show a different behavior than other 
alpha elements in super-metal-rich stars.





\section{Conclusions}

A summary of recent observational measurements of [O/Mg] versus [Mg/H] and [O/H] indicate 
that the  [O/Mg] ratio have similar trends in the Galactic bulge and solar neighborhood.


We reject the possibility that a metallicity-dependent IMF in the disk and bulge is 
responsible for the observed slopes.  The similarity of abundance trends for the two 
systems suggests a metallicity-dependent modulation of the supernova O/Mg yield ratio.  
The O/Mg slope is qualitatively consistent with a metallicity-dependent decline in oxygen 
yields due to winds from massive stars, and related to the Wolf-Rayet phenomenon, as 
suggested by McWilliam \& Rich (2004) and Fulbright et al. (2007).

To quantitatively test the significance of metallicity-dependent mass loss from massive 
stars on the O/Mg trend in the bulge and disk we have extended the bulge chemical evolution
model of Ballero et al. (2007ab) by including the metallicity-dependent oxygen yields of 
Maeder (1992) resulting from stellar winds in massive stars.  We
find that the predicted slopes and break-points of the [O/Mg] trend with [O/H] and [Mg/H],
and [O/Fe] versus [Fe/H], are very well reproduced by the enhanced model, although 
zero-point differences of up to $\sim$0.2 dex exist with observations.  The inclusion of 
the metal-dependent O yields into the Ballero et al. (2007ab) model substantially improves
the comparison between observed and predicted abundance trends.  Thus, the simple inclusion
of the known effects of mass-loss on the oxygen yields of massive stars is enough to explain 
much of the O/Mg trend in the bulge and disk.

The $\sim$0.2 dex zero-point difference between our calculations and the observations
indicates that either the semi-empirical yields of Fran\c cois et al. (2004) have
too high a sensitivity on supernova progenitor mass, or that the IMF of Ballero et al. (2007a)
is too flat; the observations might also be explained with a combined change of yields
and IMF.  Future observations and theoretical investigations into yields from massive
stars are required to resolve this issue.

By including the Maeder (1992) metal-dependent yields we remove the previous disagreement
between the IMF slope obtained by the Ballero et al. fit to the trend of [O/Fe] with [Fe/H]
compared to the slope they found by fitting the metallicity function.  Now the methods give
consistent results for the the IMF slope.


Our improvement to the chemical evolution predictions to better reproduce the O/Mg slope
greatly reduces apparent inconsistency between [O/Fe] and [Mg/Fe] trends in the high 
metallicity bulge stars: both elements are more consistent with a rapid bulge formation 
timescale, as suggested by Matteucci et al. (1999) and Ballero et al. (2007a).

A qualitative prediction of this work is that there should be an increase in carbon 
abundances for bulge stars showing the metallicity-dependent oxygen decrease; quantitative 
predictions will be made in a future investigation.  However, for carbon the comparison 
between solar neighborhood and bulge will be complicated by significant sources from low 
and intermediate mass stars in the disk.  

Finally, the metallicity-dependent oxygen yields provide a simple explanation for
the long-standing difference between the trend of [O/Fe] in the disk and that
of other alpha elements, where [O/Fe] continues to decline for stars more
metal-rich than the sun, but [Mg,Si,Ca,Ti/Fe] remain constant.

A.McWilliam acknowledges National Science Foundation grant AST~00-98612, 
including supplemental support.  F.Matteucci acknowledges support from the
Carnegie visiting scientist program.  R.M.Rich acknowledges support from
grant AST-0709479 from the National Science Foundation.




\clearpage

\clearpage



\begin{figure}
\epsscale{1.1}
\plottwo{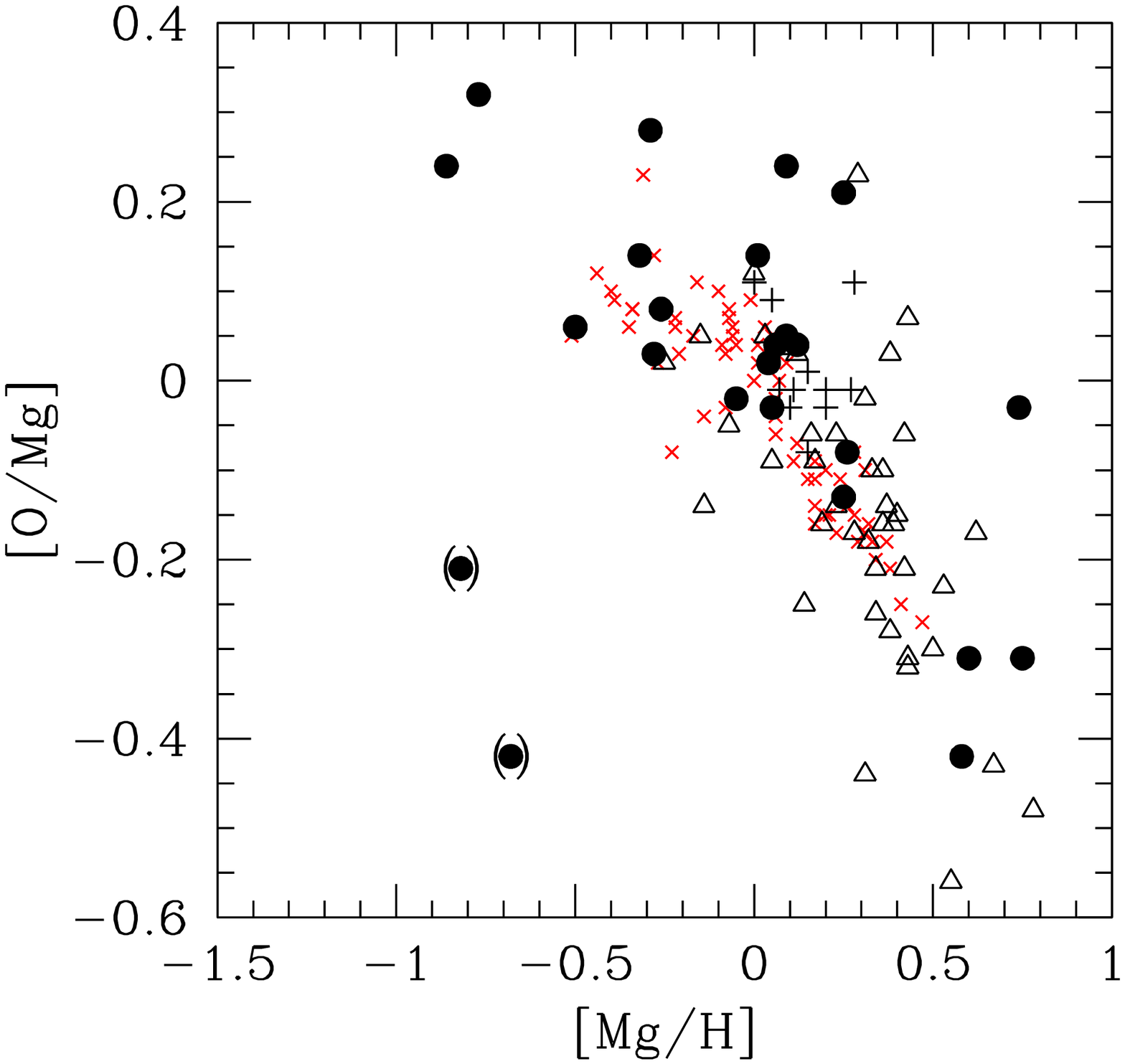}{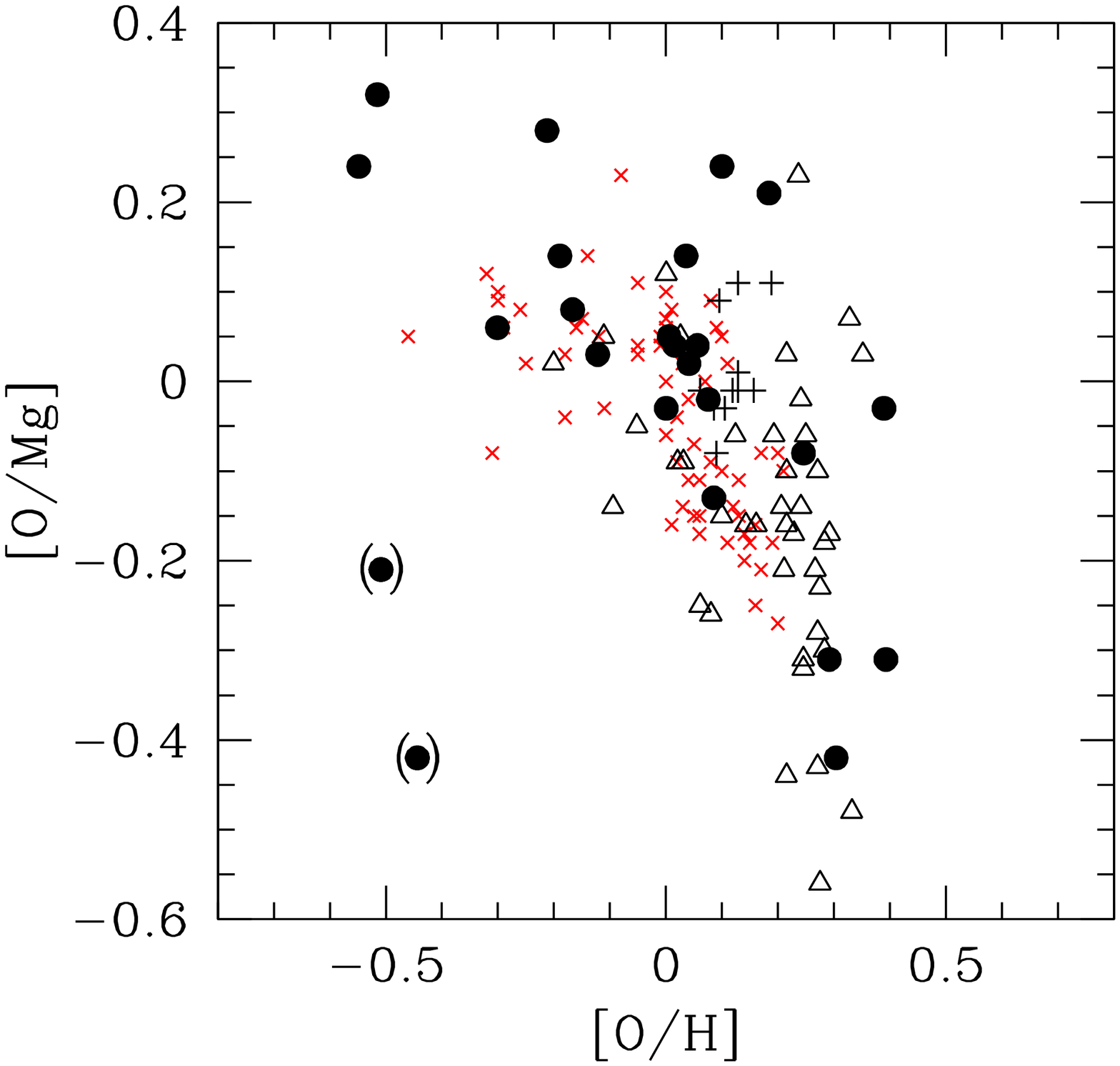}
\caption{Data for the bulge (filled circles, plus signs and open triangles) compared with
solar vicinity stars (red crosses) for [O/Mg] versus [Mg/H] (left) and [O/Mg] versus [O/H] 
(right).  In order to reduce scatter, due to the sensitivity of the measured oxygen abundance 
from a single [O~I] line, the bulge [O/H] values in the right hand figure were computed from the
[Fe/H] values and a fit to the bulge [O/Fe] versus [Fe/H] trend.  Note that the two bulge stars
in parentheses (from Fulbright et al. 2007) show the effects of proton burning products in 
their atmospheres.  Therefore, they have probably suffered a reduction in the envelope oxygen 
abundances via stellar evolution, so their oxygen abundances do not reflect the bulge composition.
\label{fig1}}
\end{figure}

\begin{figure}
\epsscale{.80}
\plotone{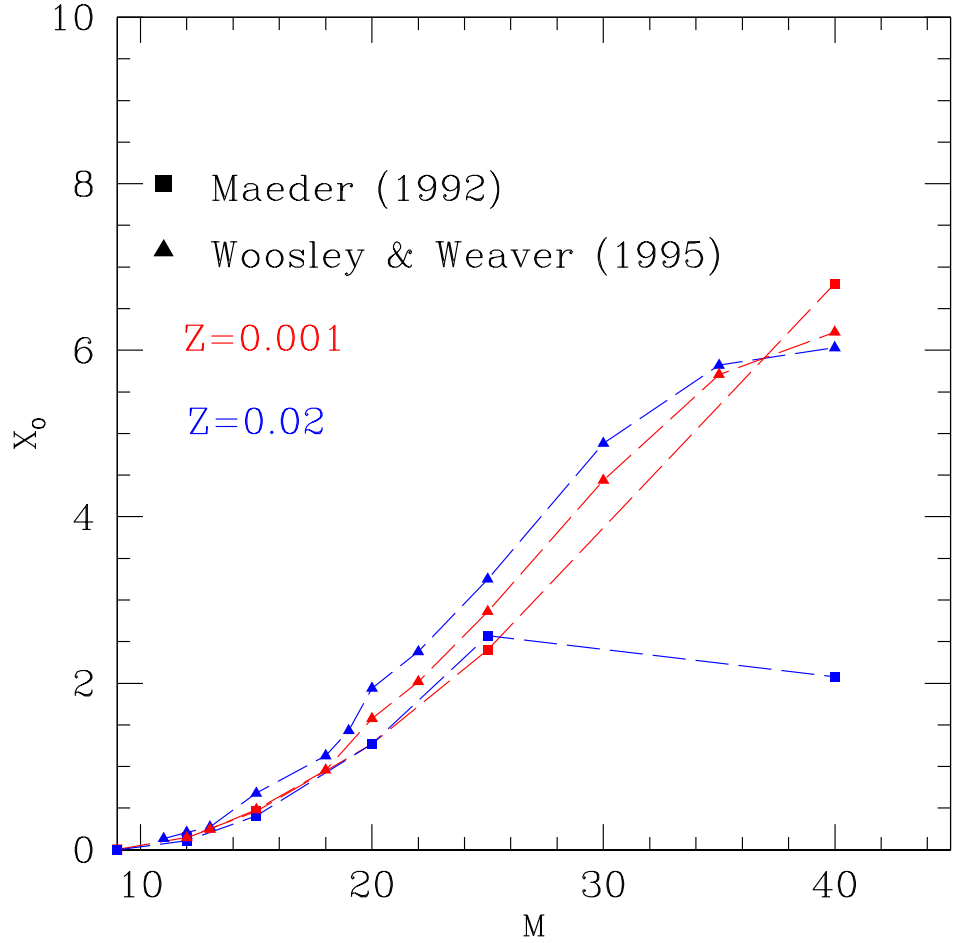}
\caption{Comparison between the O yields of Maeder (1992) and WW95 for massive stars as 
functions of initial stellar mass.\label{fig2}
The dramatic decline of O yields from high
mass stars at Solar metallicity for the Maeder
models is due to mass loss of the outer
layers, which prevents He and C from being
synthesized into O.  }
\end{figure}

\begin{figure}
\epsscale{1.1}
\plottwo{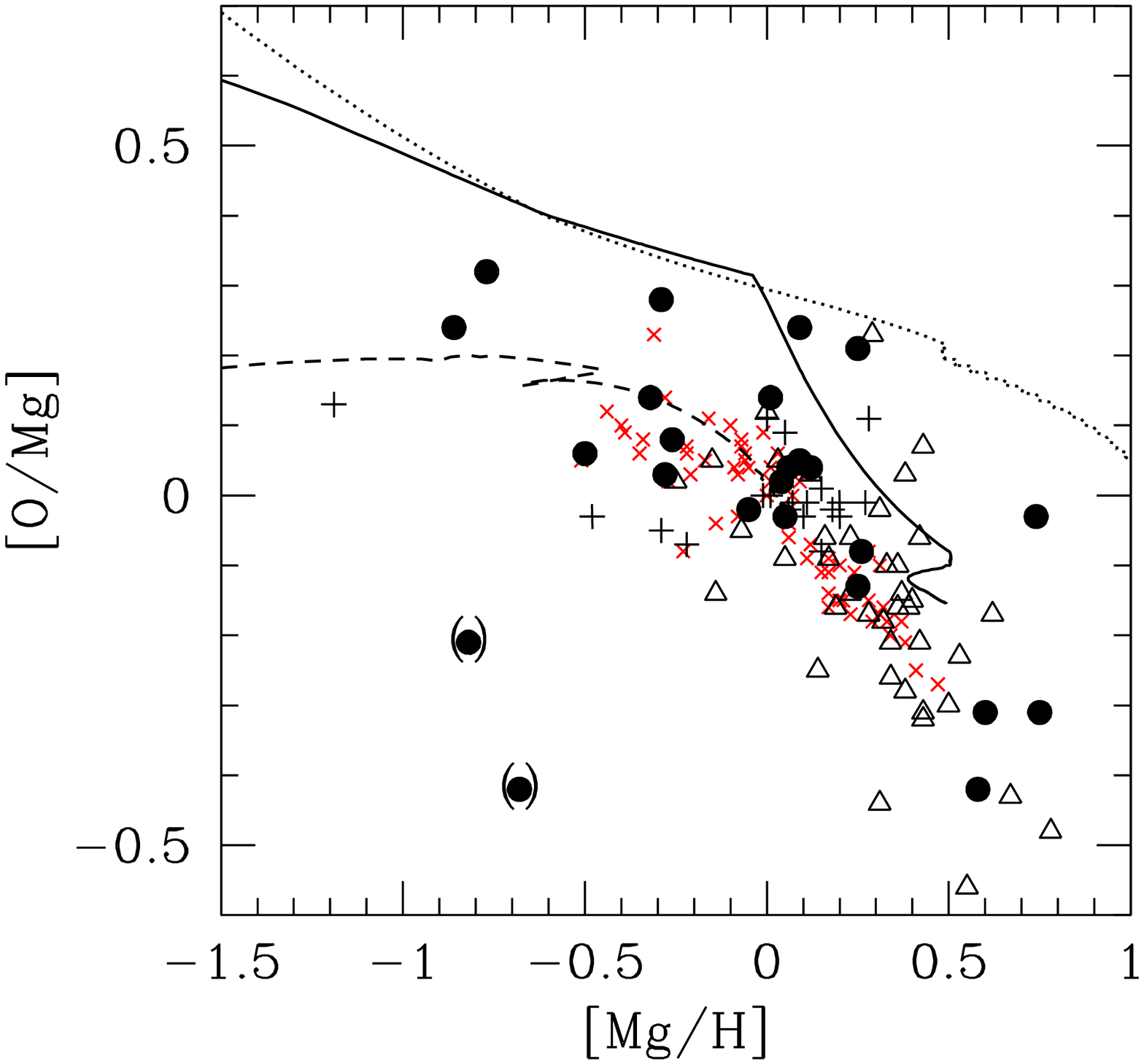}{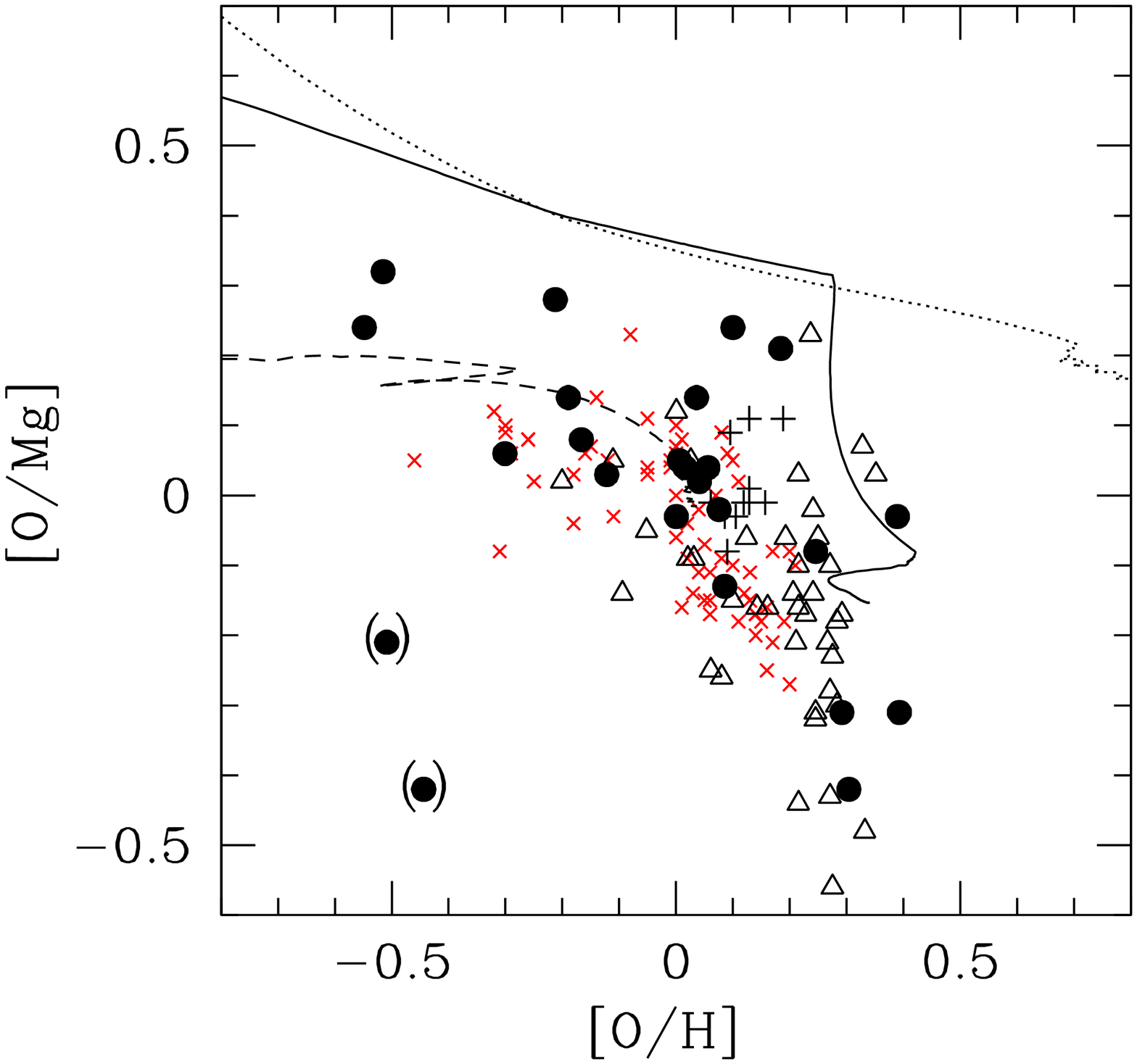}
\caption{Comparison between the predicted [O/Mg] vs. [Mg/H]  and [O/Mg] vs. [O/H] and the 
observations for the bulge and solar vicinity. The data are the same as in Fig.1. The 
continuous line is the prediction for the bulge when the Maeder (1992) O yields are considered for
metal rich massive stars.  The dotted line is the predicted [O/Mg] by Ballero et al. (2007a)
by adopting the O yields as function of metallicity by WW95.  The dashed line 
represents the prediction for the solar neighborhood when the O yields by Maeder are 
considered.  In all models we have normalized the abundances to the solar abundances as
predicted by the Milky Way model 4.5 Gyr ago. These predicted abundances are in good 
agreement with recent solar abundance determination by Asplund et al. (2005).\label{fig3}}
\end{figure}

\begin{figure}
\epsscale{.80}
\plotone{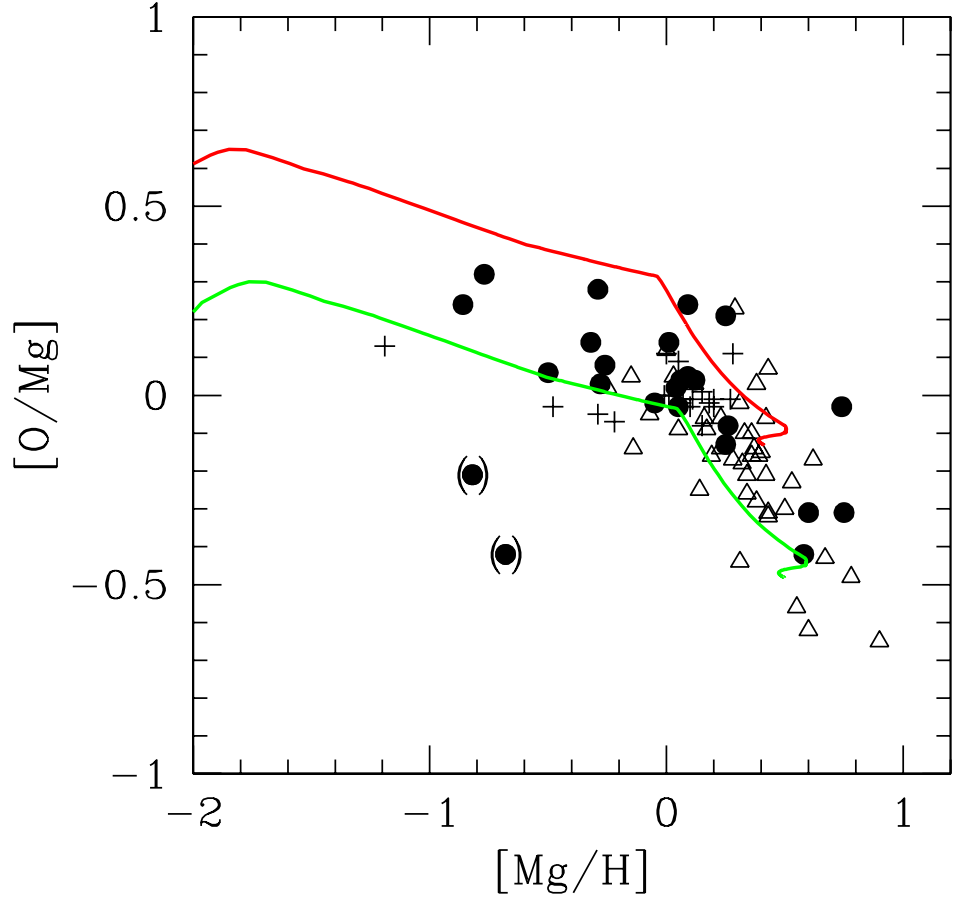}
\caption{Predicted and observed [O/Mg] vs. [Mg/H] only for the bulge.  The model (continuous line)
is the one with Maeder's yields. The upper line refers to the predicted bulge abundances
normalized to the solar predicted values by the Milky Way model, as in the curves of Figure~3.
The lower line instead refers to the predicted bulge abundances normalized Grevesse \& Sauval
(1998) solar abundances.  The offset between the two curves is mainly results from the
predicted solar abundances from our model, which is in very good agreement with the one
suggested by Asplund et al. (2005), but lower than previous estimates.  Data sources:
crosses are the infrared results from Origlia et al. (2002), Origlia \& Rich (2004), Origlia
et al. (2003), Rich \& Origlia (2005); open triangles are the data of Zoccali et al. (2006)
and Lecureur et al. (2007) (low S/N and high S/N data); filled circles are the data from 
Fulbright et al.  (2007).\label{fig4}}
\end{figure}

\begin{figure}
\epsscale{.80}
\plotone{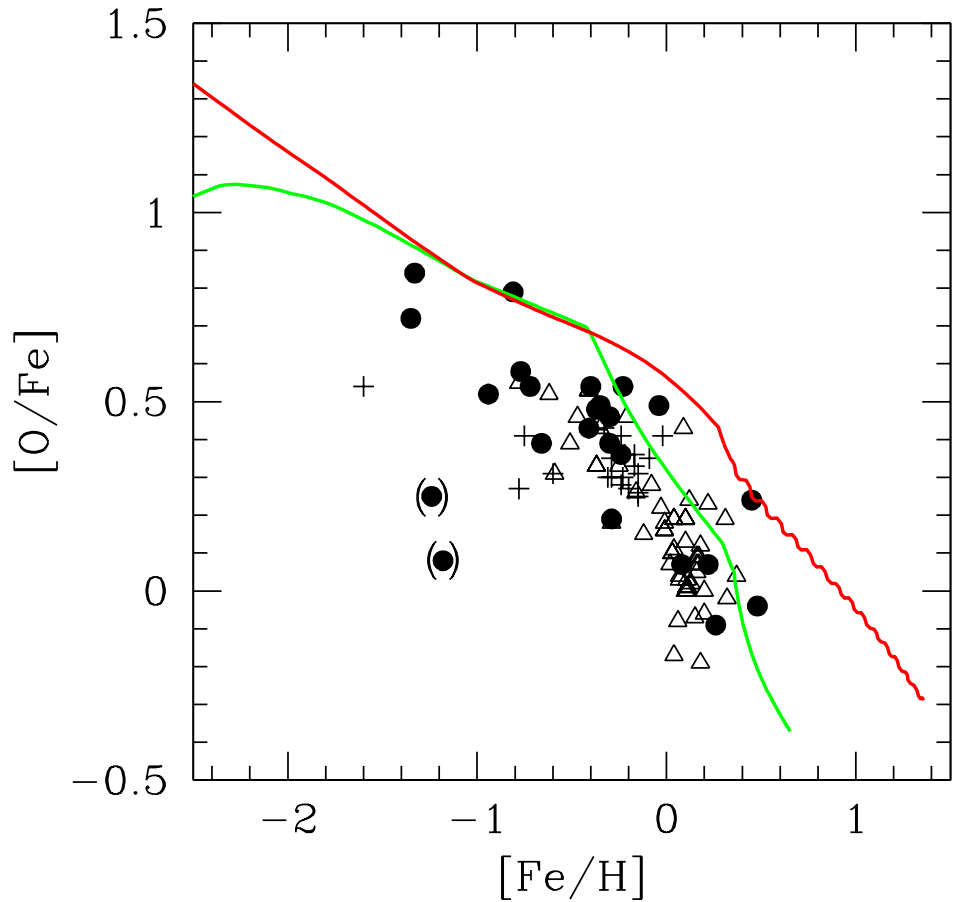}
\caption{Predicted and observed [O/Fe] vs. [Fe/H].The model with Maeder's yields is represented
 by the lower green curve whereas the predictions without mass-loss from Ballero et al. (2007a) 
are represented by the red upper line.  Both model predictions are normalized to the solar abundances 
as predicted by our Milky Way model. Data are from: Origlia \& Rich (2003; 2004), Origlia et al. 
(2003), Rich \& Origlia (2005) (crosses); Zoccali et al. (2006), Lecureur et al. (2007)
(open triangles, low S/N and high S/N data);
Fulbright et al. (2007) (filled circles).\label{fig5}}
\end{figure}

\clearpage










\clearpage







\end{document}